\documentclass[12pt]{article}
\def\theequation{\arabic{section}.\arabic{equation}}
\usepackage[centertags]{amsmath}
\usepackage{color}
\usepackage{amsfonts}
\usepackage{amssymb}
\usepackage{amsthm}
\usepackage{newlfont}

\newcounter{rown}
\def\bla{\setcounter{rown}{\value{equation}}
        \stepcounter{rown}\setcounter{equation}0
        \def\theequation{\thesection.\arabic{rown}\alph{equation}}
        }
\def\eld{\setcounter{equation}{\value{rown}}
        \def\theequation{\thesection.\arabic{equation}}}
\begin{document}

\title{
\begin{flushright}
{\small
UB-ECM-PF-08-04}\\[1.0cm]
\end{flushright}
{\bf  Enlarged NH symmetries: Particle Dynamics and Gauge
Symmetries}}

\author{\bf{Joaquim Gomis} \\
Departament d'Estructura i Constituents de la
Mat\`eria.  \\
Institut de Ciencies del Cosmos,\\ Universitat de Barcelona,
 Diagonal 647, 08028
Barcelona, Spain
\\ and \\
 {\bf Jerzy Lukierski} \\
Institute of Theoretical
 Physics, Wroc{\l}aw University, \\ 50-204 Wroc{\l}aw, Poland}

\date{}

\maketitle

\begin{abstract}
We show how the Newton-Hooke (NH) symmetries, representing a
nonrelativistic version of de-Sitter symmetries, can be enlarged by
a pair of translation vectors describing in Galilean limit the
  class of accelerations
linear in time. We study the Cartan-Maurer one-forms corresponding
to such enlarged NH symmetry group and by using cohomological
methods we determine the general 2-parameter (in $D=2+1$
4-parameter)
 central extension of
the corresponding Lie algebra. We derive by using nonlinear
realizations method the most general group - invariant particle
dynamics depending on two (in $D=2+1$ on four) central charges occurring as
the Lagrangean parameters. Due to the presence of gauge invariances
we show that for the enlarged NH symmetries quasicovariant dynamics
reduces to the one following from standard NH
 symmetries, with one central charge in arbitrary dimension $D$ and with second
 exotic central charge in $D=2+1$.
\end{abstract}

\begin{flushleft}
{\small Emails:gomis@ecm.ub.es, lukier@ift.uni.wroc.pl}
\end{flushleft}

\newpage
\section{Introduction}
In the classification of all kinematical groups in $D=3+1$
 one finds  the two nonrelativistic counterparts
of $dS$ and $AdS$ symmetries - two Newton-Hooke (NH) cosmological
groups generated
 by two NH nonrelativistic algebras \cite{Bacry:1968zf,BacNuy,Gao,GibPat}. In
comparison with the Galilei algebra, described by the generators
$J_{ij}$ (nonrelativistic rotation), $P_i$ (space translations),
$K_i$ (nonrelativistic boots) and $H$ (time translations) in NH
algebra one relation is deformed for the finite value $R$ of
nonrelativistic $(A)dS$ radius
\begin{equation}\label{golu1.1}
[H, P _i] = \pm \, \frac{K_i}{R^2} \qquad \left(
\begin{array}{ll}
+\, : \ dS  &\hbox{case} \cr - \, : \ AdS &\hbox{case}
\end{array}
\right)\, .
\end{equation}
In arbitrary dimension $D$ one can introduce one central charge
described by the mass parameter $m$
\begin{equation}\label{golu1.2}
[P_i , K_j] = m \, \delta_{ij} \, .
\end{equation}
If $D=2+1$ one can add second ``exotic" central charge $\theta$
\cite{LL,BGO,BGGK,Mariano,Gao,Alvarez:2007fw}
\begin{equation}\label{golu1.3}
[K_i, K_j ] = \theta \, \epsilon_{ij} \qquad [P_i, P_j] = -
\frac{\theta}{R^2}\, \epsilon_{ij} \, .
\end{equation}
Recently the NH symmetries were enlarged by adding the constant
accelerations, described by the generator $F_i$ \cite{golu2}. We
denote corresponding algebra by ${\widehat{\hbox{NH}}}$ and list below the
relations satisfied by new generators $F_i$, in arbitrary dimension D with central extensions included:

\bla
\begin{eqnarray}\label{golu1.4}
[J_{ij}, F_k ] &= & \delta_{ik} \, F_j - \delta_{jk} \, F_i \, ,
 \\[2pt]
 [ H , F_i ] &=& 2 K_i  \, ,
\label{golu1.4b}  \\[2pt]
[ P_i , F_j ] &=& [ F_i , F_j ] = 0 \, ,
\label{golu1.4c} \\[2pt]
 [K_i , F_j ] & = & \mp 2  m \, R^2 \, \delta_{ij}, \, \quad(-dS, + AdS).
 \label{golu1.4d}
\end{eqnarray}
We see that the relation (\ref{golu1.4d}) does not provide a finite
limit $R\to \infty$;
 in order to obtain such an Galilean limit we should introduce $R$-dependent
  Newton-Hooke  mass
 parameter
\eld
\bla
 \begin{equation}\label{golu1.5a}
m \ \longrightarrow \ m (R) = \mp \frac{c}{R^2} \, .
\end{equation}
In such a way
the relation (\ref{golu1.2}) takes the form
\begin{equation}\label{golu1.5b}
[P_i, K_j ] = \mp \frac{c}{R^2} \; \delta_{ij}
\end{equation}
\eld
and in the Galilean limit we obtain

\begin{equation}\label{golu1.6}
[ P_i , K_j] = 0 \, ,  \qquad [ K_i, F_j ] = 2c \, ,
\end{equation}
in accordance with the formulae introduced in \cite{golu3}.
Further if $D=2+1$, one can introduce two ``exotic" central charges
$\theta$, $\theta'$,
satisfying besides the relations (\ref{golu1.3}) the following
relations \cite{golu2}

\renewcommand\theequation{\arabic{section}.7\alph{equation}}
 \bla
\begin{eqnarray}
[P_i , F_j ] = 2\, \theta \, \epsilon_{ij}
\label{golu1.7a} \\[2pt]
[F_i , F_j ] = \theta' \, \epsilon_{ij} \, .
\label{golu1.7b}
\end{eqnarray}
 \eld
One of the aims of this paper is to
  enlarge the NH algebra by adding another
  vectorial Abelian generator $R_i$, and
introduce a new
 extension of  NH algebra,
 which we call doubly enlarged  $\widehat{\widehat{\hbox{NH}}}$
 algebra.
The new algebra describe the symmetries of a class of
nonrelativistic  reference frames which are characterized also by a relative
acceleration linear in time  in the flat limit
$R\longrightarrow\infty$.

 We consider a group element  of this algebra and we
    construct the Cartan-Maurer one-forms\footnote{
Some of the
   calculations with forms have being done using the Mathematica code for differential forms
   EDC \cite{bonanos}.} from which we can
calculate the nontrivial Eilenberg-Chevalley cohomology, providind
the non-trivial closed invariant two forms which  determine the
central extensions\footnote{See for example \cite{azcarragabook}.}.
In any dimension $D$
         the $\widehat{\widehat{\hbox{NH}}}$  Lie algebra has two central
         charges, and the number of central charges in $D=2+1$, similarly like for
          ${\widehat{\hbox{NH}}}$  Lie algebra \cite{golu2}, is equal to four.\footnote{We
          disregard here  in $D=2+1$ the ``trivial" central extension due to the
          deformation of the commutator $[H,J]$, where $J_{ij}=
          \epsilon_{ij}J$.}

 We further study the physical consequences of the additional
  generators $F_i$, $R_i$.
 We construct the action of a particle
    with  $\widehat{\widehat{\hbox{NH}}}$
  symmetries, by the method of non-linear
  realizations \cite{Coleman}. The action can be written as sum of WZ terms multiplied by central charge parameters.
  For a generic dimension  $D$ the lagrangrean depends on two parameters,
  but exceptionally if $D=2+1$ it depends on four parameters. The parameters
  appearing  in the lagrangean  are (besides the $dS$ radius $R$) in one to one correspondence with the
  central extensions of the $\widehat{\widehat{\hbox{NH}}}$ algebra.

The consistency of the
 field equations  imply the existence of
 one (two in $D=2+1$) relations among
 the  two (four) dimensionfull  lagrangean parameters.
These constraints permit the existence of two local gauge invariances
of the lagrangean. If a suitable gauge
 fixing is chosen the
gauge-fixed action can be identified with the one considered in
\cite{Gao,Alvarez:2007fw}. We also consider the  restriction  to the
first order $\widehat{\hbox{NH}}$ quasi lagrangean  by imposing the
vanishing of the Golstone bosons associated to generators $\vec R$.
If we impose some covariant relations representing inverse Higgs
mechanism \cite{Ivanov:1975zq} we recover the $\widehat{\hbox{NH}}$
quasi invariant higher lagrangean of reference \cite{golu2}

The plan of the paper is as follows: in section 2 we will
introduce the double enlarged $\widehat{\widehat{\hbox{NH}}}$
 algebra and its central extension; the case $D=2+1$ is considered separately.
In section 3 we construct the most general
$\widehat{\widehat{\hbox{NH}}}$ invariant first order lagrangean and
its gauge invariances. We also show that the higher order actions
considered in \cite{golu2,golu3}
 can be derived from our linear action.
  Finally in section 4 we provide some conclusions.

          \section{Doubly Enlarged Newton-Hooke Symmetries}
          \setcounter{equation}{0}

          The nonrelativistic de-Sitter and anti de-Sitter symmetries (see
          (\ref{golu1.1})) describe respectively the nonrelativistic expanding
          and nonrelativistic oscillating universes. The cosmological constant
           $\kappa = \frac{1}{R}$ describes the time scale determining the rate of
           expansion or the period of oscillations of the universe. When $\kappa
           \to 0$ we obtain the Galilean group and the standard flat nonrelativistic
           space-time.

           The Newton-Hooke algebra describing nonrelativistic (anti)-de-Sitter
           symmetries \cite{Alvarez:2007fw,golu4} was extended in \cite{golu3} to
           acceleration - extended Newton-Hooke algebra (see (\ref{golu1.1})--(\ref{golu1.6})).
           The change of nonrelativistic space-time
           ($\overrightarrow{x}= (x_1, \ldots x_d), t$)
under the transformations of
            the
           acceleration - extended NH group are the following \cite{golu3}
           \begin{eqnarray}\label{golu2.1}
\delta  x_i & = & a_i \cosh \frac{t}{R} +
 v_i \, R \sinh \frac{t}{R}
 \cr
 && + 2\, b_i \, R^2 (\cosh \frac{t}{R}  - 1) + \alpha_{ij} \, x_j
 \cr
 \delta t & = & t + a_0
           \end{eqnarray}
           with the following assignements of the parameters

\begin{tabular}{lll}
           $a_i$ &-- & spatial translations (generators $P_i$),
           dimension $\left[ a_i\right]=L^1$
\cr
           $v_i$ &-- & boosts (generators $K_i$), dimensions
           $\left[v_i\right]=L^0$
\cr
           $b_i$ &-- &  accelerations  (generators $F_i$), dimensions
           $\left[b_i\right]=L^{-1}$
\cr
           $\alpha_{ij}$ &-- &$O(d)$ space rotations (generators
           $I_{ij}$), dimensions
           $\left[\alpha_{ij}\right]=L^0$
\cr
           $a_0$ & -- &time translation (generator $H$),  dimension
           $\left[a_0\right]=L^1.$
           \end{tabular}

           Let us observe that in the flat limit $R \to \infty $ we obtain from
           (\ref{golu2.1}) the following formulae for acceleration - enlarged
           Galilean transformations
           \begin{eqnarray}\label{golu2.2}
\delta x_i & = & a_i + v_i t + b_i t^2 + \alpha_{ij} x_j \cr \delta
t & = & t  + a_0\, .
           \end{eqnarray}
           In this paper we propose to extend the formula (\ref{golu2.2}) by
           adding subsequent term
           \begin{equation}\label{golu2.3}
           \delta x_i = c_i \, t^3 \, ,
           \end{equation}
           or for $R < \infty$
           \begin{equation}\label{golu2.4}
           \delta x_i = 6 \, c_i \, R^3 (\sinh \frac{t}{R} - \frac{t}{R})
           \end{equation}
           The transformations (\ref{golu2.1}) and (\ref{golu2.4}) leads to the
           following differential
            realization of the doubly enlarged $\widehat{\widehat{\hbox{NH}}}$
           algebra
            on nonrelativistic space-time
           \begin{eqnarray}\label{golu2.5}
P_i & = & \cosh \frac{t}{R}\ \frac{\partial}{\partial x_i}
\qquad K_i = R \sinh \frac{t}{R}\ \frac{\partial}{\partial x_i}
           \cr
           F_i & = & 2 R^2 (\cosh \frac{t}{R} - 1 ) \frac{\partial}{\partial x_i}
           \cr
           R_i & = & 6 R^3  (\sinh \frac{t}{R} - \frac{t}{R}) \frac{\partial}{\partial x_i}
           \cr
           H & = &  \frac{\partial}{\partial t} \qquad
           J_{ij} = x_i  \frac{\partial}{\partial x_j}
           - x_j  \frac{\partial}{\partial x_i} \, .
\end{eqnarray}
The algebra of these vector fields give the unextended version of
the algebra $\widehat{\widehat{\hbox{NH}}}$\footnote{We consider the hyperbolic
 $dS$ case (see (\ref{golu1.1})).The trigonometric case is
 obtained by changing, $R\longrightarrow i R$ in all the equations. }

\begin{eqnarray}\label{golu2.6}
  && [H, P_i] = \frac{K_i}{R^2} \qquad [H, K_i] = P_i
  \qquad [H, F_i] = 2 K_i
\nonumber   \\[6pt]
&&   [ H, R_i ] = 3 F_i \qquad [I_A , I_B ] = 0
   \qquad I_A = (P_i, K_i, F_i, R_i)
   \nonumber   \\[6pt]
  && [I_{ij}, I_k] = \delta_{ik} I_k - \delta_{jk} I_i
   \qquad
    I_i = P_i \ \ \hbox{or} \ \ K_i \ \ \hbox{or} \ \ F_i \ \ \hbox{or}\ \ R_i
\nonumber   \\[6pt]
&& [I_{ij}, I_{kl}] = \delta_{ik} I_{jl} - \delta_{il} I_{jk} +
\delta_{jl} I_{ik} - \delta_{jk} I_{il}
\, .
\end{eqnarray}

 Let us write this Lie
algebra in terms of the dual formulation. We introduce the
  Lie-valued left-invariant Maurer-Cartan (MC)
       1-form
\begin{equation}\label{golu2.7}
         \Omega_1 = g^{-1} d g =
         L^H H + L^P_{i}P_i
         +L^{K}_{i}K_i + L^F_i F_i + L^R_l R_l + L^I_{ij}I_{ij}
         \end{equation}
where g is an general element of the group and $L^H,L^P_{i},
L^{K}_{i}, L^F_i, L^R_l,  L^I_{ij}$ are 1-forms, whose explicit form
depends on the parametrization of the group element. The MC one-form
verifies the  MC equation
       $d \Omega  + \Omega \wedge \Omega  = 0$, in terms of the
       1-forms L's we have
\begin{eqnarray}\label{golu2.8}
dL^H & = & 0 \cr dL^P_i & = & L^P_j \wedge L^I_{ij} - L^H \wedge
L^K_l \cr dL^K_i & = & L^K_i \wedge L^I_{ij} - \frac{1}{R^2}
 L^H \wedge L^P_i - 2 L^H \wedge L^F_i
 \cr
 dL^F_i & = & L^F_i \wedge L^I_{ij} - 3 L^H \wedge L^R_i
 \cr
 dL^R_i & = & L^R_i \wedge L^I_{ij}   \qquad \quad dL^I_{ij} =
 L^I_{ik} \wedge L^I_{kj} \, .
           \end{eqnarray}

If $D=2+1$ the rotations $I_{ij}$ reduce to Abelian rotation $I_{ij}
=\epsilon_{ij}I$, and the relations (\ref{golu2.8}) take the
following form:
\begin{eqnarray}\label{golu2.9}
dL^H & = & 0 \cr dL^P_i & = & \epsilon_{ij} \; L^P_j \wedge L^I -
L^H \wedge L^K_l \cr dL^K_i & = &\epsilon_{ij} \; L^K_i \wedge L^I -
\frac{1}{R^2}
 L^H \wedge L^P_i - 2 L^H \wedge L^F_i
 \cr
 dL^F_i & = & \epsilon_{ij} \;L^F_i \wedge L^I - 3 L^H \wedge L^R_i
 \cr
 dL^R_i & = & \epsilon_{ij} \; L^R_i \wedge L^I   \qquad \quad dL^I =
0\, .
           \end{eqnarray}
where $L^I_{ij} =\epsilon_{ij}L^I$.

Now we will construct the extended algebra
$\widehat{\widehat{\hbox{NH}}}$. In order to describe the central
charges in arbitrary dimension one should look for the invariant
closed 2-forms $\Omega_2$, with closure property
 $d \Omega_2 = 0$ following from the relations (\ref{golu2.8}), (\ref{golu2.9}),
 and write the 2-form, $\Omega_2=d\Omega_1$ as bilinear exterior product of the
 MC 1-forms (see e.g. \cite{azcarragabook}). One
 obtains in arbitrary dimension $D$ the following two nontrivial cohomology classes of rank 2:
 \bla
 \begin{eqnarray}\label{golu2.10a}
   \Omega^{(c)}_2 &=& L^i_{F}\wedge L^i_K - \textstyle \frac{1}{2R^2} \;
   L^i_K \wedge L^i_P + 3 L^i_P \wedge L^i_R
    \\
    \Omega^{(M)}_2  &=&  L^i_{F} \wedge L^i_R
    \label{golu2.10b}
 \end{eqnarray}

   In $D =2+1$ one obtain besides the cohomology classes (\ref{golu2.10a}--\ref{golu2.10b})
   still two ``exotic" cohomology classes, namely
\eld
\bla
 \begin{eqnarray}\label{golu2.11a}
     \widetilde{\Omega}^{(\theta)}_2
      &=&
      -
\textstyle \frac{1}{2R^2}
 \; \epsilon_{ij} \; L^P_i \wedge L^P_j
 + \textstyle \frac{1}{2} \; \epsilon_{ij}
 L^K_i \wedge L^K_j
   \nonumber    \\
      &&
      - 2 \epsilon_{ij}\;  L^F_i \wedge L^P_j
      -
      {2R^2}
      ( \epsilon_{ij} \; L^F_i \wedge L^F_j
      - 3
 \; \epsilon_{ij} \; L^K_i \wedge L^R_j)
       \\
      {\widetilde{\Omega}}^{(\theta'')}_2 &=&
\epsilon_{ij}\; L^R_i \wedge L^R_j \label{golu2.11b}
   \end{eqnarray}

In order to obtain the  Liouville one-forms $\Theta^{(c)}_1,
\Theta^{(M)}_1,
   \widetilde{\Theta}_1, \widetilde{\Theta}''_1 $
   \begin{eqnarray}\label{golu3.7}
&\Omega^{(c)}_2      = d \Theta^{(c)}_1  \qquad &\Omega^{(M)}_2  =
d\Theta^{(M)}_1 \cr
 &  \widetilde{\Omega}^{(\Theta)}_2
  = d \widetilde{\Theta}_1
\qquad &\widetilde{\Omega}^{(\Theta'')}_2
   =
d \widetilde{\Theta}''_1
   \end{eqnarray}
   one should provide a parametrization of the group element.
\eld
\bla

From the forms (\ref{golu2.10a}-b) we can construct the centrally extended Lie algebra in
arbitrary dimension $D$. The algebra is given by
$\widehat{\widehat{\hbox{NH}}}$
 relations (\ref{golu2.6}) with the commuting algebra of generators ($P_i, K_i F_i, R_i$)
  replaced by
 the non-vanishing commutation
 relations:
\begin{eqnarray}\label{golu2.12a}
[P_i, R_j ] & = & - 3 \,c \, \delta_{ij}
\\[2pt]
[K_i , P_j ] & = & \frac{1}{2 R^2}\, c \, \delta_{ij}
\label{golu2.12b}\\[2pt]
[F_i , K_j ] & = & -  c \, \delta_{ij} \label{golu2.12c}
\\[2pt]
[F_i, R_j ] & = & - M \, \delta_{ij}
 \end{eqnarray}
where c and M are two central charges.
\eld
\bla

If $D=2+1 $ from (\ref{golu2.11a}-b) we obtain two additional central charges $\theta, \theta''$:
\begin{eqnarray}\label{golu2.13a}
[F_i, F_j ] & = &  \, 4R^2 \,\theta \, \epsilon_{ij}
\\[2pt]
[P_i , P_j ] & = & \frac{1}{R^2}\, \theta \, \epsilon_{ij}
\label{golu2.13b}\\[2pt]
[F_i , P_j ] & = & 2 \, \theta \, \epsilon_{ij} \label{golu2.13c}
\\[2pt]
[K_i, K_j ] & = & -\, \theta \, \epsilon_{ij}
\label{golu2.13d}\\[2pt]
[K_i, R_j ] & = &  - 6 \,R^2 \,\theta \,
\epsilon_{ij}\label{golu2.13e}
\\[2pt]
[R_i, R_j ] & = &  -  \,\theta'' \, \epsilon_{ij}\label{golu2.13f}
 \end{eqnarray}

 We see therefore that our new enlarged
  $\widehat{\widehat{\hbox{NH}}}$ algebra in $D=2+1$
   depends on two standard  ($m$ and $M$) and two
   exotic ($\theta$ and $\theta''$) central charges.

         \section{Particle Lagrangean from Nonlinear Realizations}
          \setcounter{equation}{0}

In this section we will construct the most general particle
lagrangean (quasi) invariant under the group
$\widehat{\widehat{\hbox{NH}}}$ and its gauge invariances. We will
use the method of non-linear realizations \cite{Coleman}. As we
shall see the most general $\widehat{\widehat{\hbox{NH}}}$ first
order invariant action is endowed with gauge symmetries. We also
discuss the corresponding $\widehat{\hbox{NH}}$ first order quasi
invariant lagrangean as well as the connection with the higher order
lagrangean proposed in \cite{golu2}.

\subsection{ General Reparametrization Invariant Action}

 Let us consider the coset
of the unextended algebra
$\frac{\widehat{\widehat{\hbox{NH}}}}{SO(d)}$. We locally
parametrize the coset as follows

\renewcommand\theequation{\arabic{section} \arabic{equation}}

         \begin{equation}\label{golu3.1}
g = e^{tH} \, e^{\vec{x}\vec{P}}\, e^{\vec{v}\vec{K}}\,
e^{\vec{w}\vec{F}}\, e^{\vec{s}\vec{R}}.\end{equation}The
(Goldstone) coordinates of the coset depend on the parameter $\tau$
that parametrizes the world line of the particle (see for example
\cite{GomisKW}). The  corresponding Maurer-Cartan one-form

\renewcommand\theequation{\arabic{section}.\arabic{equation}}
         \begin{equation}\label{golu3.2}
         \Omega_1 =
         L^H H + L^P_{i}P_i
         +L^{K}_{i}K_i + L^F_i F_i + L^R_l R_l + L^I_{ij}I_{ij}
         \end{equation}
 \renewcommand\theequation{\arabic{section}.\arabic{equation}}

   In
   arbitrary dimension $D$ the 1-forms L are
   \begin{eqnarray}\label{golu3.3}
     L^H &=&  dt \cr
     L^P_i &=&  v_i\; dt + dx_i \cr
     L^K_i &=& dv_i + 2w_i dt + {\textstyle \frac{1}{R^2}} x_i dt \cr
     L^F_i &=& dw_i + 3s_i dt \cr
     L^R_i &=& ds_i\, .
   \end{eqnarray}
\bla

   Substituting the one-forms in the formulae (\ref{golu2.10a},b) and (\ref{golu2.11a},b)
   one obtains the following formulae for the corresponding Liouville one-forms,
   satisfying the relations (\ref{golu3.7})
   \begin{eqnarray}\label{golu3.9a}
     \Theta^{(c)}_1  &=& \vec{w}{ }^2 dt -
      {\textstyle \frac{1}{4R^2}} \vec{v}{ }^2 dt +
      {\textstyle
      \frac{\vec{w}\cdot \vec{x}}{R^2}} dt
    \nonumber \\[2pt]
     && +
     {\textstyle \frac{1}{4}} {\textstyle \frac{1}{(R^2)^2}}
     \vec{x}^2 dt -
     {\textstyle \frac{1}{2R^2}}
      \vec{v}\cdot d\vec{x} + \vec{w} d\vec{v}
      \nonumber \\[2pt]
      &&  - 3 \vec{s} d\vec{x} - 3 \vec{s} \cdot \vec{v} dt
     \\[2pt]
     \Theta^{(M)}_1 &=&  - \vec{s} d\vec{w} -
     {\textstyle \frac{3}{2}} \vec{s}{\; }^2 dt
     \label{golu3.9b}
   \end{eqnarray}
   and
\eld
\bla
 \begin{eqnarray}\label{golu3.5a}
     \widetilde{\Theta}_1 &=&
\epsilon_{ij}
\left(
{\textstyle \frac{1}{2}} v_i dv_j
- 2 w_i dx_j dt
+ 2 v_i w_j dt - 6 x_i s_j dt
\right)
\nonumber
\\[2pt]
      &&
+ {\textstyle \frac{\epsilon_{ij}}{R^2}}
\left(
v_i x_j dt - {\textstyle \frac{1}{2}}  x_i dx_j
\right)
      \nonumber
      \\[2pt]
      &&
- 2 \epsilon_{ij} R^2 \left(
w_i dw_j - 3 s_i dv_j - 6 s_i w_j dt
\right)
      \\[2pt]
     \widetilde{\Theta}''_1 &=&
     {\textstyle \frac{1}{2}} \; \epsilon_{ij} s_i ds_j \, .
     \label{golu3.5b}
   \end{eqnarray}
\eld

\noindent
In arbitrary dimension $D$ the action is given by
 \begin{equation}\label{golu3.12}
   S =  \int (c\Theta^{(C)}_1 + M \Theta^{(M)}_1)^*
\end{equation}
and in the  $D= 2+1$ we have
 \begin{equation}\label{golu3.13}
   \widetilde{S} = S+ \int (\theta \widetilde{\Theta}_1 +  \theta''
   \widetilde{\Theta}''_1)^*
\end{equation}
where *  means pullback on the worldline of the particle parametrized by~$\tau$.
One can fix the diffeomorphism invariance by choosing $t=\tau$. In D
dimensions the Lagrangean deduced from(\ref{golu3.12}) has the
following explicit form
\begin{eqnarray}\label{golu4.1}
  L_1 &=& c \Big[
\left(
\vec{w}^2 -
{\textstyle \frac{1}{4R^2}}
\vec{v}{\ }^2 +
{\textstyle \frac{\vec{w} \cdot \vec{x}}{R^2}} +
{\textstyle \frac{1}{4}}
\; {\textstyle \frac{1}{(R^2)^2}} \; \vec{x}{\ }^2
 - 3 \vec{s} \cdot \vec{v}
\right)
\cr
&&
 -
\left( 3\vec{s} + {\textstyle \frac{1}{2R^2}} \vec{v} \right)
{\dot{\vec{x}}} + \vec{w} \cdot {\dot{\vec{v}}} \Big] - M \left(
\vec{s}\dot{\vec w} + {\textstyle \frac{3}{2}} \vec{s}{\ }^2
\right)\, .
\end{eqnarray}
The field equations following from (\ref{golu4.1}) are
\bla
\begin{eqnarray}\label{3.14a}
\frac{1}{2R^2}
\left( {\frac{1}{R^2} } x_i + \dot{v}_i\right)
+ \frac{w_i}{R^2} + 3\dot{s}_i  = 0
\\\label{3.14b}
\frac{1}{2R^2}
\left(  \dot{x}_i + {v}_i\right)
 +\left( 3{s}_i + \dot{w}_i \right) = 0
 \\\label{3.14c}
c\left(
 { \frac{1}{R^2} } x_i + \dot{v}_i\right)
+ 2c w_i  + M \dot{s}_i = 0
\\\label{3.14d}
M\left(\dot{w}_i + 3 s_i \right)
+ 3c \left( v_i + \dot{x}_i \right) = 0 \; .
\end{eqnarray}
\eld

\subsection{Gauge Invariances and Reduction to Oscillator Dynamics}

From relations (\ref{3.14a}) and (\ref{3.14c}) follows as the consistency
condition (if $\dot{s}_i \neq 0$)
\begin{equation}\label{3.15}
M = 6 c R^2 \, .
\end{equation}
If (\ref{3.15}) is valid the pair of relations (\ref{3.14c},d) are
the same as the relations (\ref{3.14a},d). The equation
(\ref{3.14a},b) describe the following relations between the
one-forms (\ref{golu3.3}) ( $L^x= \dot{L}^x dt$, where $\dot{L}^x =
\frac{dL^x}{dt}$) \bla
\begin{eqnarray}\label{3.16a}
&&\frac{1}{2R^2} \dot{L}^P_i + \dot{L}^F_i = 0\; ,
\\
\label{3.16b}
&&\frac{1}{2R^2} \dot{L}^K_i + 3\dot{L}^R = 0\, .
\end{eqnarray}
\eld
We see that in field equations (\ref{3.16a}-b) out of four
functions $x_i (t)$, $v_i(t)$, $w_i(t)$, $s_i(t)$ two can be chosen
arbitrary. Such a property reflects the presence (if condition (\ref{3.15}) is satisfied)
of two-parameter local gauge invariance. Indeed, the action (\ref{golu4.1}) depends only
on the following two linear combinations of the field variables.
\bla
\begin{eqnarray}\label{3.17a}
u_i  & = & w_i + \frac{x_i}{2R^2} \, ,
\\
\label{3.17b}
y_i & = & s_i + \frac{1}{6R^2} v_i \, .
\end{eqnarray}
Substituting (\ref{3.17a}-b) into (\ref{golu4.1}) one obtains
\eld
\begin{equation}
L_1 = c \left(
\overrightarrow{u}^2 - 9 R^2 \overrightarrow{y}^2 - 6R^2 y_i \cdot \dot{u}_i
\right) \, .
\end{equation}
The dependence of $L$ on linear combinations  (\ref{3.17a},b) means that
 the action (\ref{golu4.1}) is invariant under the following two local gauge
  transformations, leaving the variables $u_i$ and $y_i$ invariant
  \begin{eqnarray}\label{3.19}
  \delta x_i = \epsilon_i \qquad &&
  \delta w_i = -  { \frac{1}{2}} { \frac{\epsilon_i} {R^2}}\, ,
  \cr\cr
  \delta s_i = \eta_i \qquad &&
  \delta v_i = - {\frac{1}{6}} { \frac{\eta_i} {R^2}}\; .
\end{eqnarray}
By fixing the gauge invariances (\ref{3.19})
\bla
\begin{eqnarray}\label{3.20a}
&
\epsilon_i = 2 R^2 w_i \Rightarrow\  u_i = \frac{1}{2R^2} x_i \, ,
\qquad & w_i = 0
\\
\label{3.20b}
&
\eta_i = 6 R^2 s_i \Rightarrow \ y_i = \frac{1}{6R^2} v_i \, ,
\qquad & s_i = 0
\end{eqnarray}
\eld
we obtain the following gauge-fixed form of the action (\ref{golu4.1})
\begin{equation}\label{3.21}
L^{\textrm fix}_1 = - \frac{c}{2R^2} \left( \vec{v} \dot{\vec{x}} +
\frac 12 \overrightarrow{v}{}^2 - \frac{1}2{R^2}
\overrightarrow{x}{}^2 \right) \, .
\end{equation}
The field equations derived from (\ref{3.21}) are
\begin{equation}\label{3.22}
\vec{v} + \dot{\vec{x}} = 0 \qquad \dot{\vec{v}} + \frac{1}{R^2}
\vec{x} = 0 \, ,
\end{equation}
lead to the hyperbolic oscillator equation
\begin{equation}\label{3.23}
\ddot{\vec{x}} - \frac{1}{R^2} \vec{x} = 0
\end{equation}
with the frequency described by the inverse of $dS$ radius.
\medskip

\noindent In $D=2+1$ dimensions  from (\ref{golu3.5a}-b) one obtains
the following part of the action depending on two exotic charges
$\theta, \theta''$
\begin{eqnarray}\label{3.24}
L_2  & = & \theta \Big[ {\frac{1}{2}} \epsilon_{ij} v_i \dot{v}_j -
\frac{1}{2R^2}\epsilon_{ij} x_i \dot{x}_j - 2\epsilon_{ij} w_i
\dot{x}_j + 2\epsilon_{ij} v_i w_j - 6\epsilon_{ij} x_i s_j +
\frac{\epsilon_{ij}}{R^2} v_i x_j \cr && - 2\epsilon_{ij} R^2 (w_i
\dot{w}_j - 3 s_i \dot{v}_j - 6 s_i w_j )\Big] +
\frac{\theta''}{2}\epsilon_{ij} s_i \dot{s}_j\; .
\end{eqnarray}
Introducing new variable $u_i$ (see (\ref{3.17a})) one gets

\begin{eqnarray}\label{3.25}
L_2  & = & \theta \Big( -2\epsilon_{ij} R^2 u_i \dot{u}_j + 2
\epsilon_{ij} v_i u_j + 12 \epsilon_{ij} R^2 s_i u_j \cr &&+
\frac{1}{2} \epsilon_{ij} v_i \dot{v}_j + 6 \epsilon_{ij} R^2 s_i
\dot{v}_j \Big) + \frac{\theta''}{2} \epsilon_{ij} s_i \dot{s}_j \,
.
\end{eqnarray}
the field equations are \bla
\begin{eqnarray}\label{3.25abis}
&&
2 \theta \epsilon_{ij} \left(
2R^2 \dot{u}_j + v_j + 6 R^2 s_j
\right)
= 0
\\
&& \epsilon_{ij} \left( \dot{v}_j + 2u_j + 6R^2 \dot{s}_j \right) =
0
\\
&& 6 \theta R^2 \epsilon_{ij} \left( \dot{v}_j + 2 u_j \right) +
\theta'' \epsilon_{ij} \dot{s}_j = 0
\end{eqnarray}
\eld by consistency  (if $\dot{s}_i  \neq 0$) we have
\begin{equation}\label{3.26}
\theta'' = 36 R^4 \theta \; .
\end{equation}
In such a case the action (\ref{3.25}) depends only on two variables (\ref{3.17a}-b)
\begin{equation}\label{3.27}
L_2 = \theta R^2
\Big(-2 \epsilon_{ij} u_i \dot{u}_j
+ 12 \epsilon_{ij} y_i u_j + 18 \epsilon_{ij}
R^2 y_i \dot{y}_j
\Big) \;.
\end{equation}

The most general action in $D=2+1$ is described by the sum of the
action (\ref{golu4.1}) and (\ref{3.24}).
 If we write down the field equations it appears that one can derive the following
 equation
 \begin{equation}\label{3.27bis}
A \dot{s}_i + B \epsilon_{ij} \dot{s}_j = 0
 \end{equation}
where
\begin{equation}\label{3.27bbis}
A = M - 6 cR^2 \qquad \quad B = \theta'' - 36 R^4 \theta \; .
\end{equation}
The solvability of Eq. (\ref{3.27bis}) leads to
\begin{equation}\label{3.27bbbis}
\det \left( \begin{array}{cc} A & B \cr -B & A \end{array}
\right) = 0 \quad \Rightarrow \ A^2 + B^2 = 0 \quad
\Rightarrow \ A = 0, \ B = 0
\end{equation}
i.e. we reproduce the constraints (\ref{3.15}) and (\ref{3.26}).

Using the conditions (\ref{3.20a}-b) and (\ref{3.15}), (\ref{3.26})
  one obtains the following general
gauge-fixed action in $D=2+1$
\begin{eqnarray}\label{3.28}
L^{\textrm fix} & = & L^{\textrm fix}_1 +
L^{\textrm fix}_2 =
 - \frac{c}{4R^2}
 \Big(
 2v_i \dot{x}_i + \overrightarrow{v}^2 - \frac{1}{R^2} \overrightarrow{x}^2
 \Big)
 \cr
 && +
 \theta \epsilon_{ij}\frac 12\Big(
 - \frac{x_i \dot{x}_j}{R^2}
 - 2\frac{x_i v_j}{R^2} +
    v_i \dot{v}_j
 \Big)
\end{eqnarray}
which can be identified with the one presented in
\cite{Alvarez:2007fw}
 (see formula (\ref{golu3.5a},b)) if we pass from $dS$ to $AdS$ case
 ($R \to i R$), parametrize velocity  with opposite sign
  ($v_i \to - v_i $) and put $m = - \frac{c}{2R^2}$, $\theta = \frac{1}{2}
  \kappa$.
  One can state therefore that the physical content of our model is the same as in the case
  of standard Newton-Hooke symmetries, with the same number of generators as in Galilean case.

\subsection{Relation with $\widehat{\hbox{NH}}$ Higher  Order Lagrangean}

Firstly we derive the $\widehat{\hbox{NH}}$  first order lagrangean
from our results presented above. For that purpose we should put
equal to zero the Goldstone fields associated with generators $\vec
R$ , ($\vec s=0$). For dimension $D\neq 2+1$ the action density
becomes $L_1|_{\vec s=0}$, where the lagrangean $L_1$ is given by
(\ref {golu4.1}). Instead for D=2+1 there is  an extra term
$L_2|_{\vec s=0}$ with $L_2$ described (\ref{3.24}).

Let us use the variables
\begin{eqnarray}\label{newtrans}
u_i  & = & w_i + \frac{x_i}{2R^2} \, ,
\\
 y_i & = & \frac{1}{6R^2} v_i \, .
\end{eqnarray}

 We can write the Lagrangeans as
 \begin{eqnarray}
L_1|_{\vec s=0}&=& \tilde{c} \left( 2 \overrightarrow{u}^2 - \frac
{1} {2R^2} \overrightarrow{v}^2 - 2 v_i \cdot \dot{u}_i \right)
\label{gl33}\\
L_2|_{\vec s=0}&=& \theta' \Big(\epsilon_{ij} u_i \dot{u}_j - \frac
{1} {2R^2} \epsilon_{ij} v_i u_j -  \frac {1} {4R^2} \epsilon_{ij}
v_i \dot{y}_j \Big)
\label{gl331}\;.
\end{eqnarray}

where $\tilde c=\frac c 2, \theta'=-4R^2\theta$.

If we employ the covariant inverse Higgs mechanism
\cite{Ivanov:1975zq}, by putting $L_i^K=L_i^P=0$ and using (\ref
{golu3.3}), (\ref{newtrans})  we obtain
\begin{eqnarray}
 v_i &=& - \dot x_i
\\
 u_i&=& -\frac 12 \dot v_i\,.
\end{eqnarray}
From  (\ref{gl33}) and  (\ref{gl331}) we get the
$\widehat{\hbox{NH}}$ higher order Lagrangean proposed in
\cite{golu2}.

 \section{Conclusions}
   \setcounter{equation}{0}
 We demonstrated that the action constructed via Cartan-Maurer one-forms
   for extended NH groups with additional
   constant  and linear  acceleration generators ($F_i, R_i$)
   supplements only
    a gauge sector of  the standard Newton-Hooke dynamics. To
    achieve such equivalence we have to reduce the number of central
charges via
    suitably chosen relations  (\ref{3.15}) and (\ref{3.25}).
    We recall that changing the sign of $R^2$   (see (\ref{golu1.1}))
    we obtain from the hyperbolic oscillator (see (\ref{3.23}))
     the standard one, with trigonometric solutions
     (see \cite{Alvarez:2007fw}).

     It seems therefore that if consider more general NH algebras with
     extra Abelian generators, the particle dynamics will be
     equivalent to the one describing ordinary NH particle. This is due the
     appearance of gauge symmetries associated the extra generators.
     It will be interesting to understand better such a property.

We have also seen that if we assume that  Goldstone fields
associated with $\vec R$ vanish and we impose some covariant
relations one recovers the $\widehat{\hbox{NH}}$ higher order
lagrangians introduced in reference \cite{golu2}.

Finally we would like to observe that the transformations
(\ref{golu2.1})-(\ref{golu2.4}), for $R\longrightarrow\infty$, can
be extended as follows
\begin{equation}
x^{'}_i=x_i+a_i(t)\,.
\label{ex}
\end{equation}
We get in such a way the reparametrization group of Newton's
equation of particle in the presence of nonrelativistic
gravitational potential $V(x,t)$(see e.g. \cite{golufb}).

\begin{equation}
\frac{d^2x_i}{dt^2}=\frac{\partial V}{\partial x_i}
\label{exsys}
\end{equation}
which is invariant under (\ref{ex}) if $V'=V-\ddot{a}_ix_i$. It
seems interesting to consider the link between the extension of
system (\ref{exsys}) for R finite and the enlarged NH symmetries of
arbitrary order. \vskip 3mm

\section*{Acknowledgements}
We acknowledge Roberto Casalbuoni, Gary Gibbons, Mikhail Plyushchay,
Peter Stichel and Andrzej Trautmann for useful discussions. This
work has been supported by the European EC-RTN project
MRTN-CT-2004-005104, MCYT FPA 2007-66665, CIRIT GC 2005SGR-00564,
Spanish Consolider-Ingenio 2010 Programme CPAN (CSD2007-00042) and
by Polish MNiSW grant N N202 318534 (JL). We would like to thank the
PH-TH Division at CERN for its hospitality where this works has
started.

\end{document}